\documentclass[journal,12pt,onecolumn,draftclsnofoot]{IEEEtran}

\usepackage{pgfplots}
\usepackage[bookmarks,colorlinks]{hyperref}
\usepackage[linesnumbered,ruled,lined]{algorithm2e}
\usepackage{enumitem}
\usepackage{algpseudocode}
\usetikzlibrary{shapes.multipart,intersections}
\usepackage{cite}
\usepackage{amsmath,amssymb,amsfonts,amsthm,steinmetz}
\usepackage{graphicx}
\usepackage{mathrsfs}  
\usepackage{textcomp}
\usepackage{acronym}
\usepackage{xcolor}
\usepackage{upgreek,xspace}

\usepackage{tikz}
\usetikzlibrary{calc}
\makeatletter
\newcommand{\gettikzxy}[3]{%
  \tikz@scan@one@point\pgfutil@firstofone#1\relax
  \edef#2{\the\pgf@x}%
  \edef#3{\the\pgf@y}%
}
\makeatother

\usepackage[draft]{todonotes}

\usepackage{esvect}

\usepackage{times}
\usepackage{bm}
\usepackage{amsmath}
\usepackage{amssymb}
\usepackage{stmaryrd}
\usepackage{babel}
\usepackage{graphics, graphicx}
\usepackage{xcolor}
\usepackage{gensymb}
\usepackage{cite}
\usepackage{enumitem}
\usepackage{url}

\begin{document}

\title{Efficient Computation of Physics-Compliant Channel Realizations for (Rich-Scattering) RIS-Parametrized Radio Environments}

%\title{Computationally Efficient Use of Physics-Compliant Channel Models for Arbitrarily Complex RIS-Parametrized Radio Environments}

\author{Hugo~Prod'homme~and~Philipp~del~Hougne,~\IEEEmembership{Member,~IEEE}
\thanks{P.d.H. acknowledges funding from the CNRS pr\'{e}maturation program (project ``MetaFilt'') and the ANR PRCI program (project ANR-22-CE93-0010-01).
}
\thanks{
H.~Prod'homme and P.~del~Hougne are with Univ Rennes, CNRS, IETR - UMR 6164, F-35000, Rennes, France (e-mail: \{hugo.prodhomme; philipp.del-hougne\}@univ-rennes1.fr).
}
\thanks{\textit{(Corresponding Author: Philipp del Hougne.)}}
}

\maketitle

\begin{abstract}
Physics-compliant channel models of RIS-parametrized radio environments require the inversion of an ``interaction matrix'' to capture the mutual coupling between wireless entities (transmitters, receivers, RIS, environmental scattering objects) due to proximity \textit{and} reverberation. 
The computational cost of this matrix inversion is typically dictated by the environmental scattering objects in non-trivial radio environments, and scales unfavorably with the latter's complexity.
In addition, many problems of interest in wireless communications (RIS optimization, fast fading, object or user-equipment localization, etc.) require the computation of multiple channel realizations. 
To overcome the potentially prohibitive computational cost of using physics-compliant channel models, we \textit{i)} introduce an isospectral reduction of the interaction matrix from the canonical basis to an equivalent reduced basis of primary wireless entities (antennas and RIS), and \textit{ii)} leverage the fact that interaction matrices for different channel realizations only differ regarding RIS configurations and/or some wireless entities’ locations.
\end{abstract}


\section{Introduction}\label{sec_Introduction}

%  computational complexity of physics-compliant models risks to thwart their proliferation and use

Smart radio environments in which reconfigurable intelligent surfaces (RISs) endow wireless system engineers with the ability to control the wireless channel (in addition to the usual ability to control the transmitted signals) are considered a paradigm shift that may impact future wireless network generations. However, the modeling of RIS-parametrized wireless channels is still in its infancy and wide-spread cascaded channel models tacitly assume that multi-bounce paths can be neglected~\cite{rabault2023tacit}. At the same time, the computational cost of existing physics-compliant models (e.g., PhysFad~\cite{PhysFad}) can rapidly become excessive under rich-scattering conditions, especially if multiple channel realizations are required. 
The computation of multiple channel realizations is required in common problems involving RIS optimization, wireless localization and sensing, and/or fast fading, because a generic physics-compliant wireless channel depends non-linearly on the RIS configuration and/or on the location of the wireless entities~\cite{rabault2023tacit,ChloeMag}.
Here, we reduce the potentially prohibitive computational cost of evaluating multiple physics-compliant realizations of generic RIS-parametrized wireless channels by several orders of magnitude through the introduction of a reduced-basis representation of the wireless system and through efficient use of the knowledge about relations between different realizations that usually only differ regarding the RIS configuration or the locations of some wireless entities.

A RIS-parametrized wireless channel is a \textit{linear} input-output relation that, in general, depends \textit{non-linearly} on the RIS configuration due to \textit{i)} proximity-induced mutual coupling between neighboring RIS elements, and \textit{ii)} reverberation-induced long-range coupling between all RIS elements~\cite{rabault2023tacit}. Physically, this ``structural non-linearity'' originates from paths involving multiple bounces between RIS elements [for \textit{i)}] and between the RIS and other wireless entities (antennas and scattering environment) [for \textit{ii)}]~\cite{rabault2023tacit}. 
Mathematically, this ``structural non-linearity'' manifests itself in physics-compliant models via the inversion of an ``interaction matrix'' that can be cast in terms of infinite matrix power series~\cite{rabault2023tacit}. Wide-spread cascaded models tacitly assume that these infinite series can be truncated early on such that all paths involving more than one encounter with a RIS element are neglected~\cite{rabault2023tacit}.

Recently, one of the current authors (and coworkers) introduced a physics-compliant end-to-end channel model for arbitrarily complex RIS-parametrized radio environments derived from first physical principles: PhysFad~\cite{PhysFad}. 
Since the RIS-parametrized radio environment is a linear time-invariant\footnote{The time invariance is to be understood with respect to the scale of the wave period. A given channel realization always corresponds to a fixed (not time-varying) RIS configuration.} electrodynamical system, there must be a linear operator describing the link between the incident electromagnetic fields and the polarization fields they induce in the system. Assuming a sufficiently-high-resolution discretization of the system into polarizable elements, this operator is proportional to the inverse of an ``interaction matrix''; the latter's $i$th diagonal entry is the inverse polarizability of the $i$th polarizable element and its $(i,j)$th off-diagonal entry is the free-space Green's function between the $i$th and $j$th polarizable elements. The polarization fields depend hence non-locally on the incident electromagnetic field because of the coupling between different polarizable elements via the non-zero off-diagonal entries of the interaction matrix. 
For simplicity of notation and exposition, PhysFad~\cite{PhysFad} describes each wireless entity (antenna, RIS element, scattering object) as a dipole or collection of dipoles.\footnote{Extensions to multi-pole expansions of each polarizable element are conceptually straight-forward.} The scattering response of any arbitrary anisotropic object can be equivalently described by a finite collection of fictitious dipoles~\cite{bertrand2020global}. A procedure to model an RIS prototype with such a discrete-dipole framework is detailed and validated in Ref.\cite{diebold2023reflectarray}.

Equivalent formulations of PhysFad for the strongly limiting restriction of the radio environment being free space were presented in terms of impedance matrices in Refs.\cite{williams2020communication,gradoni_EndtoEnd_2020,badheka2023accurate,akrout2023physically} and in terms of scattering parameter network analysis in Refs.~\cite{shen2021modeling,franek2023electromagnetics} (of which Refs.~\cite{williams2020communication,gradoni_EndtoEnd_2020,shen2021modeling} appeared prior to PhysFad). The impedance-matrix-based formulations build on an earlier proposal of a multiport circuit theory of communications systems~\cite{ivrlavc2010toward,ivrlavc2014multiport} that took mutual coupling between the elements of antenna arrays into account, but did not consider the possibility of RIS-parametrized wave propagation environments (including potentially highly complex scattering structures). Very recently, Ref.~\cite{mursia2023modeling} reproduced the PhysFad formalism in terms of impedance matrices for radio environments involving scattering objects.

Important potential deployment scenarios of RISs at microwave and millimeter-wave frequencies, which are a significant component of 6G’s all-spectra-integrated networks~\cite{you2021towards}, are confronted with rich scattering within the radio environment~\cite{GeorgeMag}. A prototypical example is RIS-assisted machine-type communication in factories~\cite{giordani2020toward}. 
The required size of the interaction matrix to accurately describe such rich-scattering radio environments can be very large, implying a potentially prohibitively large computational cost if multiple channel realizations must be computed.

In this Letter, we alleviate this computational cost by several orders of magnitude. To this end, on the one hand, we leverage an isospectral reduction of the interaction matrix to an equivalent reduced basis of primary wireless entities (antennas and RIS). Thereby, we interpret the scattering between primary and secondary wireless entities (scattering objects) as additional coupling mechanisms between the primary wireless entities.
One of the current authors (and coworkers) recently used a similar reduced-basis representation to achieve covert scattering control in metamaterials with non-locally encoded hidden symmetries~\cite{HiddenSymmetry}. The reduced-basis approach presented in this Letter in the context of RIS-parametrized radio environments can be straighforwardly transposed to the physics-compliant modelling of dynamic metasurface antennas~\cite{yoo2019analytic,williams2022electromagnetic}.
On the other hand, we leverage the insight that different channel realizations typically only differ regarding some of the diagonal entries and/or some rows and columns of the interaction matrix such that updating a previously evaluated wireless channel is computationally much cheaper than evaluating it from scratch.

This Letter is organized as follows. In Sec.~\ref{sec_Generalities}, we briefly review the PhysFad formalism. In Sec.~\ref{sec_ReducedBasisRepresentation}, we introduce the reduced-basis representation of PhysFad. 
In Sec.~\ref{sec_opportunities}, we explain how previous channel realizations can be updated to account for a new RIS configuration (Sec.~\ref{subsec_RISconfiguration}), a change of the scattering objects' properties to sweep the importance of multi-path propagation (Sec.~\ref{sec_KQfactor}), the displacement of a wireless entity such as the user equipment or scattering objects (Sec.~\ref{subsec_MovingUE}), or any combination thereof (Sec.~\ref{subsec_combinedUpdate}).
We close with a brief conclusion %and outlook
in Sec.~\ref{sec_Conclusion}.

\textit{Notation.} The vector $\mathbf{a}$ containing the diagonal entries of the matrix $\mathbf{A}$ is denoted by $\mathbf{a} = \mathrm{diag}(\mathbf{A})$.

\section{Generalities}\label{sec_Generalities}

Within the PhysFad framework outlined above and detailed in Ref.~\cite{PhysFad}, the $i$th dipole, located at position $\mathbf{r_i}$, has a frequency-dependent polarizability $\alpha_i$ that relates the induced dipole moment $p_i$ to the incident electromagnetic field $E_i$.\footnote{For simplicity of exposition and notation, we describe a 2D setting but a dyadic extension to the more general 3D case is straight-forward.} The induced dipole moment will reradiate an electromagnetic field whose strength at the location $\mathbf{r_j}$ of the $j$th dipole is $G_{ji}p_i$, where $G_{ji}$ is the free-space Green's function between positions $\mathbf{r_i}$ and $\mathbf{r_j}$. The simplest model of a RIS element is a dipole with Lorentzian polarizability whose resonance frequency is reconfigured upon changing the RIS element's configuration~\cite{PhysFad}. Thereby, the intertwinement of phase and amplitude response and their frequency selectivity are automatically captured. 

Each wireless entity [T (transmitter), R (receiver), S (RIS), or E (scattering environment)] is described as a dipole or collection of dipoles~\cite{PhysFad}, yielding a total number of $N=N_{\mathrm{T}}+N_{\mathrm{R}}+N_{\mathrm{S}}+N_{\mathrm{E}}$ dipoles. 
The diagonal entries of the ``interaction matrix'' $\mathbf{W} \in \mathbb{C}^{N\times N}$ are the dipoles' inverse polarizabilities, and the off-diagonal entries of $\mathbf{W}$ are the negatives of the corresponding Green's functions:
\begin{equation}
W_{i,j}=\begin{cases}
\alpha_{i}^{-1}, & i=j\\
-G_{ij}, & i\neq j
\end{cases}.
\end{equation}
By grouping the dipoles according to which wireless entity they belong to, $\mathbf{W}$ can be written in block form:
\begin{equation}
\mathbf{W}=\begin{bmatrix}
\mathbf{W}_{\mathrm{TT}} & \mathbf{W}_{\mathrm{TR}} & \mathbf{W}_{\mathrm{TS}} & \mathbf{W}_{\mathrm{TE}}\\
\mathbf{W}_{\mathrm{RT}} & \mathbf{W}_{\mathrm{RR}} & \mathbf{W}_{\mathrm{RS}} & \mathbf{W}_{\mathrm{RE}}\\
\mathbf{W}_{\mathrm{ST}} & \mathbf{W}_{\mathrm{SR}} & \mathbf{W}_{\mathrm{SS}} & \mathbf{W}_{\mathrm{SE}}\\
\mathbf{W}_{\mathrm{ET}} & \mathbf{W}_{\mathrm{ER}} & \mathbf{W}_{\mathrm{ES}} & \mathbf{W}_{\mathrm{EE}}
\end{bmatrix}.
\end{equation}
The RIS configuration $\mathbf{c} = [ \alpha_{N_{\mathrm{T}}+N_{\mathrm{R}}+1}^{-1} \dots \alpha_{N_{\mathrm{T}}+N_{\mathrm{R}}+N_{\mathrm{S}}}^{-1} ] \in\mathbb{C}^{N_{\mathrm{S}}}$ is encoded in the $\mathbf{W}_\mathrm{SS}$ block: $\mathrm{diag}(\mathbf{W}_\mathrm{SS}) = \mathbf{c}$. 
The physics-compliant end-to-end channel matrix $\mathbf{H}\in\mathbb{C}^{N_{\mathrm{R}}\times N_{\mathrm{T}}}$ (assuming identical transmitting and receiving antennas  for simplicity) is proportional to the RT block of the inverse of $\mathbf{W}$:
\begin{equation}
    \mathbf{H} \propto [\mathbf{W}^{-1}]_{\mathrm{RT}}.
\label{eq:channeldef}
\end{equation}
The non-linear dependence of $\mathbf{H}$ on $\mathbf{c}$ is hence apparent.

\section{Reduced-Basis Representation}\label{sec_ReducedBasisRepresentation}

\begin{figure}[h]
\centering
\includegraphics[width=0.6\linewidth]{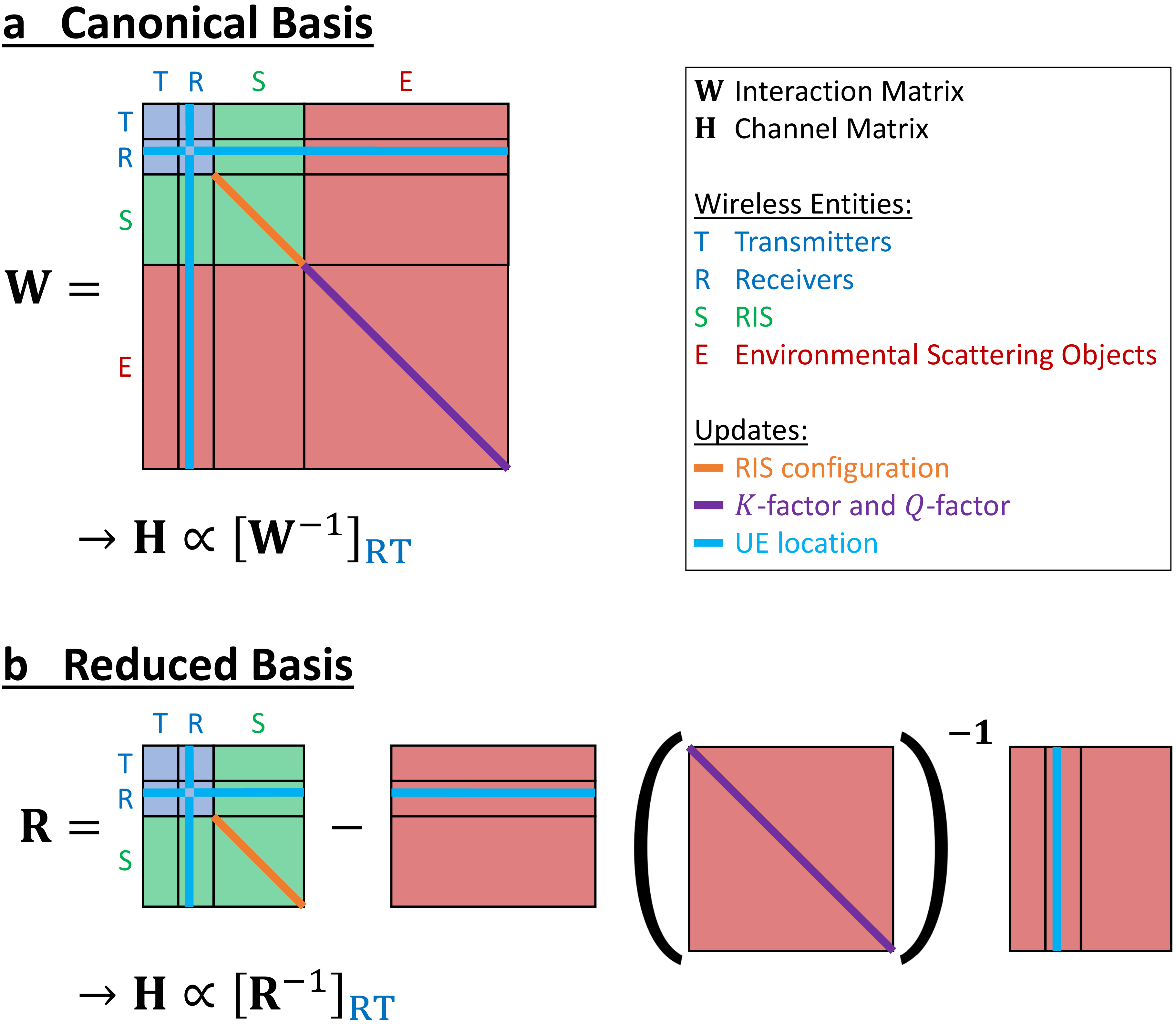}
\caption{Schematic illustration of  a physics-compliant model's interaction matrix in the canonical (a) and reduced (b) basis. In addition, the parts of the interaction matrix affected by three types of updates are highlighted. }
\label{fig1}
\end{figure}

The wireless system engineer only has direct access to a subset $\mathcal{P}$ of the wireless system's $N$ internal scattering entities, namely to the $N_{\mathrm{T}}+N_{\mathrm{R}}$ antennas (for injecting and capturing waves) and the $N_{\mathrm{S}}$ RIS elements (for configuring the RIS). 
Hence, we seek an \textit{equivalent} representation of the wireless system reduced to the subset $\mathcal{P}$ of the $p=N_{\mathrm{T}}+N_{\mathrm{R}}+N_{\mathrm{S}}$ ``primary'' dipoles from the groups T, R and S.
We straight-forwardly rewrite $\mathbf{W}$ as a $2 \times 2$ block matrix:
\begin{equation}
    \mathbf{W} = \begin{bmatrix} 
	\mathbf{W}_{\mathcal{PP}}  & \mathbf{W}_{\mathcal{P\Bar{P}}} \\
	\mathbf{W}_{\mathcal{\Bar{P}P}}  & \mathbf{W}_{\mathcal{\Bar{P}\Bar{P}}} \\
\end{bmatrix},
\label{eq:reducedbasis}
\end{equation}

\noindent where $\mathcal{\Bar{P}}$ denotes the subset of the $s=N-p=N_{\mathrm{E}}$ ``secondary'' dipoles from the group E. Under rich-scattering conditions, usually $p \ll s $ because many more dipoles are necessary to describe the scattering environment than to describe the antennas and the RIS.
As illustrated in Fig.~\ref{fig1}, standard formulas for the inversion of a block matrix yield
\begin{equation}
    [\mathbf{W}^{-1}]_{\mathcal{PP}} = \left( \mathbf{W}_{\mathcal{PP}} - \mathbf{W}_{\mathcal{P\Bar{P}}} \mathbf{W}_{\mathcal{\Bar{P}\Bar{P}}}^{-1} \mathbf{W}_{\mathcal{\Bar{P}P}} \right)^{-1} = \mathbf{R}^{-1}
\label{eq:blockwiseinversion}
\end{equation}
and Eq.~(\ref{eq:channeldef}) is hence equivalent to $\mathbf{H} \propto [\mathbf{R}^{-1}]_\mathrm{RT}$.

The dimensionality of the interaction matrix underpinning physics-compliant channel models of non-trivial RIS-parametrized radio environments involving scattering objects can hence be significantly reduced by operating in an equivalent reduced basis of primary wireless entities as opposed to the usual canonical basis. In this reduced representation, the ($i,j$)th off-diagonal entry of $\mathbf{R}$ accounts for coupling between the $i$th and $j$th RIS element due to proximity \textit{and} reverberation, whereas the ($i,j$)th off-diagonal entry of $\mathbf{W}$ only accounts for proximity-induced coupling. In addition, reverberation adds a self-coupling term such that the diagonals of $\mathbf{W}$ and $\mathbf{R}$ differ, too.

\section{Computing Multiple Channel Realizations}\label{sec_opportunities}

The interaction matrices underlying different channel realizations often only differ in some details such that it is more efficient to update a previously computed channel realization than to evaluate the new channel realization from scratch.
Three types of such updates of particular interest are highlighted in Fig.~\ref{fig1} and treated in the subsequent sub-sections:

\begin{enumerate}[label=(\roman*)]
    \item A change in RIS configuration corresponds to a change of the diagonal of $\mathbf{W}_{\mathrm{SS}}$ (highlighted in orange in Fig.~\ref{fig1}a).
    \item A change in the location of the $i$th dipole corresponds to a change of the $i$th row and the $i$th column of $\mathbf{W}$, excluding the $i$th diagonal entry (highlighted in cyan in Fig.~\ref{fig1}a). This case is relevant to problems in wireless localization (of a moving user equipment and/or a moving non-cooperative scattering object) as well as fast-fading scenarios (the locations of some environmental scattering objects differ across different realizations).
    \item A change in the properties of the environmental scattering objects corresponds to a change of the diagonal of $\mathbf{W}_{\mathrm{EE}}$ (highlighted in purple in Fig.~\ref{fig1}a). Such a change is required in order to sweep the level of reverberation (quantified, e.g., by the $Q$-factor~\cite{west2017best}), and it is also required in combination with (ii) to sweep through different $K$-factors (see Sec.~III-F~in~Ref.\cite{PhysFad}).

\end{enumerate}

Regarding the quantification of computational complexity of matrix operations, the inversion of an $n \times n$ matrix will be considered to have a complexity of $\mathcal{O}\left(n^3\right)$ and the product of an $n_1 \times n_2$ matrix with an $n_2 \times n_3$ matrix will be considered to have a complexity of $\mathcal{O}\left(n_1 n_2 n_3\right)$ in the following.

\subsection{Updating the RIS configuration}\label{subsec_RISconfiguration}

A modification of the configuration of $m \leq N_{\mathrm{S}}$ RIS elements implies an update of $m$ diagonal entries of $\mathbf{W}_{\mathrm{SS}}$.\footnote{The same method could also be applied if there was a change of the transmitting and/or receiving antennas' polarizabilities.} The Woodbury matrix identity~\cite{hager1989updating} allows us to obtain the updated values of $\mathbf{W}^{-1}$ upon expressing the modification $\mathbf{\Delta W}$ of $\mathbf{W}$ as a product of three matrices, $\mathbf{\Delta W}=\mathbf{UCV}$,
\begin{equation}
    \mathbf{\left(W+\Delta W\right)}^{-1}=\mathbf{\left(W+UCV\right)}^{-1} =\mathbf{W}^{-1}-\mathbf{W}^{-1}\mathbf{U}\left(\mathbf{C}^{-1}+\mathbf{VW}^{-1}\mathbf{U}\right)^{-1}\mathbf{VW}^{-1}.
\label{eq:woodbury}
\end{equation}
Here, $\mathbf{C} \in \mathbb{C}^{m\times m}$ is a diagonal matrix containing the changes of the inverse polarizabilities of the $m$ modified RIS elements: $\mathrm{diag}(\mathbf{C}) = [ \Delta\alpha_{n_1}^{-1} \dots \Delta\alpha_{n_m}^{-1} ]$, where $n_k$ is the index of the $k$th modified RIS element (the modified RIS elements are not necessarily contiguous). The matrices $\mathbf{U} \in \mathbb{B}^{N\times m}$ and $\mathbf{V} \in \mathbb{B}^{m\times N}$ are constructed from unit vectors and related through a transpose operation. They act as selectors through matrix multiplication such that the computation of their dot products with $\mathbf{W}^{-1}$ can be replaced by a simple row-selection or column-selection operation. 
%The columns and rows selected are the ones corresponding to the modified elements. 
With $\delta_{i,j}$ denoting the Kronecker delta, $\mathbf{C}$, $\mathbf{U}$ and $\mathbf{V}$ can be expressed as follows:
\begin{equation}
C_{k,k'}=\delta_{k,k'}\Delta\alpha_{n_k}^{-1}, \ \ \ 
U_{i,k} = V_{k,i} = \delta_{i,n_k}.
\end{equation}

This method applies to generic continuously tunable RIS elements, i.e., $\alpha_{n_k}^{-1}$ can take any physically plausible complex value. Many experimental prototypes have 1-bit programmable RIS elements (e.g., Refs.~\cite{rabault2023tacit,GeorgeMag}). In these cases, the inverse polarizabilities of the RIS elements can only take two possible complex values, such that it is more efficient to apply Eq.~(\ref{eq:woodbury}) by considering as starting point an interaction matrix $\mathbf{W}$ corresponding to one out of two complementary RIS configurations. Thereby, the configuration of at most $m=\lfloor N_\mathrm{S}/2 \rfloor$ RIS elements must be updated which reduces the computational cost of complexity $\mathcal{O}\left(m^3\right)$ for inverting the $m \times m$ matrix sum in Eq.~(\ref{eq:woodbury}). The two remaining matrix products (after the selection operations with $\mathbf{U}$ and $\mathbf{V}$) have a computational complexity of $\mathcal{O}\left(Nm^2\right)$.

The same approach can be applied in the reduced basis with the advantage of having to store only the $p\times p$ matrix $\mathbf{R}^{-1}$ rather than the $N \times N$ matrix $\mathbf{W}^{-1}$ in memory.
Since the computation of $\mathbf{H}$ only requires the evaluation of the RT block of $\mathbf{R}^{-1}$ (or $\mathbf{W}^{-1}$), the complete evaluation of $\mathbf{\left(W+UCV\right)}^{-1}$ in Eq.~(\ref{eq:woodbury}) is not required. The computationally most efficient approach is
\begin{equation}
[\left(\mathbf{R}+\mathbf{UCV}\right)^{-1}]_{\mathrm{RT}} = [\mathbf{R}^{-1}]_{\mathrm{RT}} 
- \left[\mathbf{R}^{-1}\mathbf{U}\right]_{\mathrm{R:}}\left(\mathbf{C}^{-1}+\mathbf{VR}^{-1}\mathbf{U}\right)^{-1}\left[\mathbf{V}\mathbf{R}^{-1}\right]_{\mathrm{:T}},
\label{eq:reducedwoodbury}
\end{equation}
where the sub-index $_\mathrm{[RT]}$ denotes the selection of the rows and columns corresponding to the groups R and T,  the sub-index $_\mathrm{[R:]}$ denotes the selection of the rows corresponding to group R (and all the columns), and the sub-index $_\mathrm{[:T]}$ denotes the selection of the columns corresponding to group T (and all the rows).
The computational complexity of the inner matrix inversion in Eq.~(\ref{eq:reducedwoodbury}) is still $\mathcal{O}\left(m^3\right)$, but it is $\mathcal{O}\left(N_\mathrm{R}m^2\right)$ for the rightmost matrix product and $\mathcal{O}\left(N_\mathrm{T}m^2\right)$ for the leftmost matrix product.

\textit{Remark\,:} The inversion of a matrix $\mathbf{A}$ is actually implemented by solving $\mathbf{Ax}=\mathbf{I}$, where $\mathbf{I}$ is the identity matrix. Thus, a product of the form $\mathbf{A}^{-1}\mathbf{B}$ is computed faster by solving $\mathbf{Ax}=\mathbf{B}$ than by inverting $\mathbf{A}$ and multiplying by $\mathbf{B}$ afterwards. This insight allows for a further reduction of the computational cost of Eq.~(\ref{eq:reducedwoodbury}).

\subsection{Updating the $K$ and/or $Q$ factor}
\label{sec_KQfactor}

The amount of reverberation (multipath propagation) inside the radio environment is determined by the properties of the dipoles from the group E. Changing the amount of reverberation directly alters the radio environment's $Q$-factor. In addition, if fast fading is implemented by moving a fixed subset of the dipoles of the group E, then changing their properties will also alter the $K$-factor (see Sec.~III-F in Ref.~\cite{PhysFad}). The ability to update the properties of the dipoles from the group E is hence essential to sweep through different types of radio environments, from free space to rich scattering. Such changes require (within the PhysFad use case considered in Sec.~III-F of Ref.~\cite{PhysFad}) an identical update of the identical $s = N_{\mathrm{E}}$ diagonal entries of $\mathbf{W}_{\mathcal{\Bar{P}\Bar{P}}} = \mathbf{W}_{\mathrm{EE}}$.

According to Eq.~(\ref{eq:blockwiseinversion}), the full evaluation of $\mathbf{W}_{\mathcal{\Bar{P}\Bar{P}}}^{-1}$ is required to obtain the wireless channel. To efficiently update this inversion upon an identical change of all diagonal entries of $\mathbf{W}_{\mathcal{\Bar{P}\Bar{P}}}$, we use the eigendecomposition $\mathbf{W}_{\mathcal{\Bar{P}\Bar{P}}} = \mathbf{Q_{\mathcal{\Bar{P}\Bar{P}}} D_{\mathcal{\Bar{P}\Bar{P}}}}\mathbf{Q_{\mathcal{\Bar{P}\Bar{P}}}}^{-1}$, where $\mathbf{D_{\mathcal{\Bar{P}\Bar{P}}}}$ is a diagonal matrix whose diagonal entries are the eigenvalues of $\mathbf{W}_{\mathcal{\Bar{P}\Bar{P}}}$ and the columns of $\mathbf{Q_{\mathcal{\Bar{P}\Bar{P}}}}$ contain the corresponding eigenvectors. 
The eigendecomposition after subtracting the same complex-valued scalar $\lambda$ from all diagonal
entries is
\begin{equation}
\left(\mathbf{W}_{\mathcal{\Bar{P}\Bar{P}}}-\lambda \mathbf{I}\right)^{-1} = \mathbf{Q}_{\mathcal{\Bar{P}\Bar{P}}}\left(\mathbf{D}_{\mathcal{\Bar{P}\Bar{P}}}-\lambda \mathbf{I}_s\right)^{-1} \mathbf{Q}_{\mathcal{\Bar{P}\Bar{P}}}^{-1}.
\label{eq:eigenupdate}
\end{equation}

\noindent Inserting Eq.~(\ref{eq:eigenupdate}) into Eq.~(\ref{eq:blockwiseinversion}) yields
\begin{equation}
        \mathbf{R}^{-1}_{\lambda} = 
        \left(\mathbf{W}_{\mathcal{PP}} - 
        \mathbf{W}_{\mathcal{P\Bar{P}}} \mathbf{Q}_{\mathcal{\Bar{P}\Bar{P}}}
        \left(\mathbf{D}_{\mathcal{\Bar{P}\Bar{P}}}-\lambda \mathbf{I}_s\right)^{-1} 
        \mathbf{Q}_{\mathcal{\Bar{P}\Bar{P}}}^{-1} \mathbf{W}_{\mathcal{\Bar{P}P}} 
        \right)^{-1},
\label{eq:blockwiseeigenupdate}
\end{equation}
where $\mathbf{R}^{-1}_{\lambda}$ denotes the updated version of $\mathbf{R}^{-1}$.
The products $\mathbf{\Sigma}=\mathbf{W}_{\mathcal{P\Bar{P}}} \mathbf{Q}_{\mathcal{\Bar{P}\Bar{P}}}$ and $\mathbf{\Psi}=\mathbf{Q}_{\mathcal{\Bar{P}\Bar{P}}}^{-1} \mathbf{W}_{\mathcal{\Bar{P}P}}$ of dimensions $p \times s$ and $s \times p$, respectively, can be pre-computed and stored. We can then rewrite Eq.~(\ref{eq:blockwiseeigenupdate}) in the reduced basis representation as 
\begin{equation}
        \mathbf{R}^{-1}_{\lambda} = 
        \left( \mathbf{W}_{\mathcal{PP}} - \mathbf{\Sigma} \left(\mathbf{D}_{\mathcal{\Bar{P}\Bar{P}}}-\lambda \mathbf{I}_s\right)^{-1} \mathbf{\Psi}
        \right)^{-1}.
\label{eq:reducedeigenupdate}
\end{equation}

The evaluation of Eq.~(\ref{eq:reducedeigenupdate}) does not require access to $\mathbf{Q}_{\mathcal{\Bar{P}\Bar{P}}}$ or $\mathbf{Q}_{\mathcal{\Bar{P}\Bar{P}}}^{-1}$ once the smaller matrices $\mathbf{\Sigma}$ and $\mathbf{\Psi}$ are pre-computed. Moreover, the inversion of the diagonal matrix $\left(\mathbf{D}_{\mathcal{\Bar{P}\Bar{P}}}-\lambda \mathbf{I}_s\right)$ can be performed analytically. In addition, one of the matrix products with the diagonal matrix in Eq.~(\ref{eq:reducedeigenupdate}) can be replaced with an element-wise multiplication of a row with a matrix. The remaining matrix product has a computational complexity of $\mathcal{O}\left(sp^2\right)$. The remaining outer matrix inversion has a computational complexity of $\mathcal{O}\left(p^3\right)$. This constitutes a computationally very efficient way to obtain the update $\mathbf{R}^{-1}_{\lambda}$, as the $\mathcal{O}\left(s^3\right)$ computational cost of the inversion of $\mathbf{W}_{\mathcal{\Bar{P}\Bar{P}}}$ is skipped.

\subsection{Updating the location of a wireless entity}\label{subsec_MovingUE}

A change in the location of the $j$th dipole implies that all off-diagonal entries of the $j$th row and the $j$th column of $\mathbf{W}$ are altered. The Woodbury matrix identity from Eq.~(\ref{eq:woodbury}) can be applied to this rank-2 update of $\mathbf{W}$ to obtain the corresponding updated $\mathbf{W}^{-1}$.
To this end, we choose
\begin{equation}
\mathbf{C}=\mathbf{I}_2,\ \ \ 
\mathbf{U}=\left[\begin{array}{cc} \mathbf{\Delta G_{j}}^{\mathrm{T}} & \boldsymbol{\delta}_\mathbf{j}^{\mathrm{T}}\end{array}\right],\ \ \ 
\mathbf{V}=\left[\begin{array}{c} \boldsymbol{\delta}_\mathbf{j} \\ \mathbf{\Delta G_{j}}\end{array}\right],
\end{equation}
where $\mathbf{\Delta G_{j}}=[\Delta G_{j,1} \dots \Delta G_{j,N}] \in \mathbb{C}^{N}$ contains the changes of the Green's functions between all $N$ dipoles and the displaced
dipole indexed $j$ (with $\Delta G_{j,j}=0$), and $\boldsymbol{\delta}_\mathbf{j}=[\delta_{j,1} \dots \delta_{j,N}]  \in \mathbb{B}^{N}$ is a unit vector.
The computational cost of evaluating Eq.~(\ref{eq:woodbury}) in this case is dominated by the three matrix products involving $\mathbf{W}^{-1}$ which have a computational complexity of $\mathcal{O}\left(N^2\right)$. 
$\mathbf{W}^{-1}$ must be stored in memory in its entirety to evaluate $\mathbf{VW}^{-1}\mathbf{U}$.

In order to apply this approach in the reduced basis, the displaced dipole must be a primary dipole. In the case of a moving user equipment (receiver), this is naturally the case. In the case of moving scattering objects, we must divide the group E into dynamic and static dipoles, and include the former within $\mathcal{P}$ rather than $\mathcal{\Bar{P}}$. Indeed, a wireless channel in a RIS-parametrized dynamic radio environment is subject to a non-linear double-parametrization through both the RIS and the dynamic scattering objects~\cite{ChloeMag}.

Assuming the displaced dipole is contained within $\mathcal{P}$, the values of the $\mathbf{W}$ sub-matrices to be updated are one row of $\mathbf{W}_\mathcal{P\Bar{P}}$, one column of $\mathbf{W}_\mathcal{\Bar{P}P}$, and the off-diagonal entries of one row and one column of $\mathbf{W}_\mathcal{PP}$, as illustrated by the cyan-colored line in Fig.~\ref{fig1}b. The computation of the two matrix products within the central term of Eq.~(\ref{eq:blockwiseinversion}) involves the $\mathbf{W}_\mathcal{\Bar{P}\Bar{P}}$ matrix, and both have a computational complexity of $\mathcal{O}\left(ps^2\right)$.

If all possible positions of the displaced dipole are known (e.g., in the case of a scattering object moving along a known trajectory such as a robot in a factory), it is furthermore possible to pre-compute all possible values of the row of $\mathbf{W}_\mathcal{P\Bar{P}}$ corresponding to the displaced dipole. Due to reciprocity ($\mathbf{W}$ is symmetric), the corresponding column of $\mathbf{W}_\mathcal{\Bar{P}P}$ is simply the transpose of this row of $\mathbf{W}_\mathcal{P\Bar{P}}$. It is then possible to avoid the need for computing the inverse of $\mathbf{W}_\mathcal{\Bar{P}\Bar{P}}$ by using Eq.~(\ref{eq:reducedeigenupdate}) with $\lambda=0$. The previously identified updated row of $\mathbf{W}_\mathcal{P\Bar{P}}$ and corresponding column of $\mathbf{W}_\mathcal{\Bar{P}P}$ translate into an updated row of $\mathbf{\Sigma}$ and an updated column of $\mathbf{\Psi}$. 
These possible row and column values are not related through a transpose operation. They can be pre-computed using $\mathbf{\Sigma}=\mathbf{W}_{\mathcal{P\Bar{P}}} \mathbf{Q}_{\mathcal{\Bar{P}\Bar{P}}}$ and $\mathbf{\Psi}=\mathbf{Q}_{\mathcal{\Bar{P}\Bar{P}}}^{-1} \mathbf{W}_{\mathcal{\Bar{P}P}}$ as in the previous subsection. Once this pre-computation is completed, there is no need to store $\mathbf{Q}_{\Bar{P}\Bar{P}}$ and $\mathbf{Q}_{\Bar{P}\Bar{P}}^{-1}$ (nor $\mathbf{W}_\mathcal{P\Bar{P}}$ or $\mathbf{W}_\mathcal{\Bar{P}P}$) in memory and the displacement of a dipole can be taken into account by setting the corresponding row of $\mathbf{\Sigma}$ and column of $\mathbf{\Psi}$ to the corresponding pre-computed values, as well as setting the corresponding off-diagonal row and column of $\mathbf{W}_\mathcal{PP}$ to the corresponding values, and then evaluating Eq.~(\ref{eq:reducedeigenupdate}).

If multiple dipoles are displaced simultaneously, as many rows of $\mathbf{\Sigma}$, columns of $\mathbf{\Psi}$, and off-diagonal row-column couples of $\mathbf{W}_\mathcal{PP}$ need to be updated as there are displaced dipoles. As the updated rows of $\mathbf{\Sigma}$ and columns of $\mathbf{\Psi}$ describe the interaction between one moving dipole and all the secondary static dipoles, the possible values are independent of any other displaced dipoles. In contrast, the off-diagonal terms in $\mathbf{W}_\mathcal{PP}$ describing the Green's functions between two moving dipoles obviously depend on the location of both involved moving dipoles, and hence, for the sake of implementation simplicity, should be re-computed for every realization rather than being pre-computed.

\subsection{Combined updates}
\label{subsec_combinedUpdate}

The three update methods described in the previous three sections are mutually compatible and can hence be arbitrarily combined. 
For example, the resulting updated matrix $\mathbf{R}^{-1}_{\lambda}$ from Sec.~\ref{sec_KQfactor} or Sec.~\ref{subsec_MovingUE} can be used within Eq.~(\ref{eq:reducedwoodbury}) to generate channel realizations for different RIS configurations. Moreover, simultaneous updates of both the inverse polarizabilities of the secondary dipoles and the location of some primary dipoles are possible.

\section{Conclusion}\label{sec_Conclusion}

In order to alleviate the prohibitively large computational cost of evaluating multiple physics-compliant realizations of RIS-parametrized (rich-scattering) channels, we have introduced \textit{i)} a reduced-basis representation of the underlying interaction matrix, and \textit{ii)} identified efficient ways to update a previous channel realization based on how it differs from a new required realization in terms of diagonal and/or off-diagonal modifications of the underlying interaction matrix.
Thereby, we enable orders-of-magnitude improvements of the  computational cost of using physics-compliant models in problems involving RIS optimization, wireless localization and/or fast fading.
Moreover, our analysis provides new physical insights into the inner working of physics-compliant channel models of generic RIS-parametrized wireless channels.

\bibliographystyle{IEEEtran}
%\bibliography{references}


\end{document}